\documentclass[twoside,openright]{nictatechreport}
\usepackage[lined,boxed,commentsnumbered]{algorithm2e}
\usepackage{graphicx}
\usepackage{url}
\usepackage{subfig} 

\title{Can SDN Mitigate Disasters?}
\author{Vincent Gramoli,\textsuperscript{$^\star$,1,2} Guillaume Jourjon,\textsuperscript1 Olivier Mehani\textsuperscript1
} 

\newcommand{\theKeywords}{SDN, distributed system }

\affiliation{$^\star$Corresponding author: \email{vincent.gramoli@sydney.edu.au}\\
\textsuperscript1 Nicta, Sydney. Eveleigh, NSW, Australia, \email{first.last@nicta.com.au}\\
\textsuperscript2 University of Sydney, Sydney, NSW, Australia, \email{vincent.gramoli@sydney.edu.au}}
\date{}


\reportnumber{XXXX}
\copyrightyear{2014}
\headertitle{Gramoli {\em et al.} Can SDN Mitigate Disasters?}

\begin{document}

\begin{abstract}
Datacenter networks and services are at risk in the face of disasters. Existing fault-tolerant 
storage services cannot even achieve a nil recovery point objective (RPO) as client-generated 
data may get lost before the termination of their migration across geo-replicated datacenters.
SDN has proved instrumental in exploiting application-level information
to optimise the routing of information.
In this paper, we propose \emph{Software Defined Edge (SDE)} or the implementation of SDN 
at the network edge to achieve nil RPO. We illustrate our proposal with a 
fault-tolerant key-value store that experimentally recovers from disaster within 30s.
Although SDE is inherently fault-tolerant and scalable, 
its deployment raises new challenges on the partnership between ISPs and CDN providers.
We conclude that failure detection information at the SDN-level can effectively benefit applications
to recover from disaster.
\end{abstract}


\frontmatter

\mainmatter
\section{Introduction}

With the advent of cloud services, the computation needed by individuals gets progressively centralised at few datacenters around the world. 
This centralisation puts the services at risk in the face of disasters, like nuclear power plant explosions or earthquakes, that
can affect a large geographical region.
This risk motivated, several years ago, some Wall Street financial institutions to build
datacenters outside the blast radius of a nuclear attack on New York City, hence drawing a ring of land 
in New Jersey called the ``Donut''.\footnote{http://www.datacenterknowledge.com/archives/2008/03/10/new-york-donut-boosts-nj-data-centers/}
Within a range of 30 to 70 km from the city centre, these backup datacenters aim at maintaining synchronous data transfer with the city centre 
while providing service continuity despite disasters.

This idea as well as all actions taken to ensure service continuity in the event of a disaster are called \emph{disaster recovery}~\cite{GPH90}.
Recovering some services can be particularly challenging.
Take a fault-tolerant key-value store, which serves \texttt{get} and \texttt{put} requests from clients,  
as an example. To guarantee that data persist despite the failure of isolated servers, 
the effect of an update, say a successful \texttt{put} request, should at least be replicated at multiple servers of a datacenter.
Such a fault-tolerant approach is however not sufficient to recover the data upon a disaster, as all servers of this datacenter could be affected.
Prevalent solutions investigated asynchronous migration and mirroring techniques across 
datacenters of distant geographical locations~\cite{PMF+02,WLR+11}.
The challenge is then to minimise the migration delay as this always translates into some amount of data that can be lost during a disaster, also known as 
\emph{Recovery Point Objective (RPO)}.

While industrials have made remarkable efforts in reaching low RPO, a disaster often leads to network outage,
preventing remote clients from accessing the running backup service during a large delay, referred to as the \emph{Recovery Time Objective (RTO)}.
More specifically, if a backup server starts rapidly operating using a different IP address, it may not be instantaneously accessible as refreshing the 
DNS table at the edge of the network could take hours or even days.
The main problem is that the network itself takes a long time for recovering from a disaster, hence delaying the application recovery.
Actually, introducing disaster recovery capabilities at the network level should intuitively help achieving both nil RTO and RPO.
The challenge lies in distributing the network control that is usually centralised in the network core---typically in the datacenter network---to the network edge.

In this paper, we propose a solution to disaster recovery at the level of the network edge by exploiting \emph{Software Defined Network (SDN)}~\cite{MAB+08}, namely the 
decoupling of control functions from the data processing and forwarding functions remotely controllable.	
With the recent advancement in switch fabrics, 
major content delivery network providers started rerouting the traffic depending on congestion and type of network services. 
An example is the use of SDN technology by Google for speeding up video streaming content
as part of the YouTube service~\cite{Vah13}. 
SDN technology is already available in off-the-shelf network equipment from
vendors such as CISCO, Juniper or NEC.
More specifically, our solution is to adapt routers at the network edge (1) to prepare for a disaster by proactively replicating the traffic of ``critical'' data towards multiple datacenters 
and (2) to cope with a disaster by reactively redirecting the traffic from a failed datacenter to another datacenter located in a non impacted region.

We illustrate our solution by deploying a distributed key-value store application on top of an emulation of our SDN edge or \emph{Software Defined Edge (SDE)} for short.
A distributed key-value store is a popular storage abstraction offered by many cloud service providers as part of MongoDB, IBM's Spinnaker, Amazon's SimpleDB, Cassandra or 
WindowsAzure Table Storage. 
The popularity of this storage abstraction is in part due to its simplicity in associating a key to a value for simple storage and retrieval of data.
Our key-value store is strongly consistent and tolerates isolated failures by exploiting intra-datacenter replication but relies on our SDE solution to cope with disasters.
This application is an ideal use-case to illustrate that SDE can reach a nil RPO and a 30s RTO by using 
TCP retransmissions directly at the router level upon the detection of edge failure and until traffic redirection between datacenter regions as far as Australia to Ireland.

Our SDE solution is not a panacea but raises new kinds of challenges in the deployment of SDN technologies at large scale.
An interesting observation is that implementing SDE seems to require a tight collaboration between content 
delivery network providers and ISPs so that the latter offer an API that the 
former could exploit to control forwarding rules and mitigate disasters.
We are not aware of any such partnership and there is certainly a long way before SDE could be deployed at the scale of Internet.
We also believe however that there already exist incentives for such a collaboration. In particular, we conjecture
that SDE would reduce the operational cost of network forensics of these companies as unidirectional traffic such as SYN packets 
would no longer occurs towards offline datacenter, and thus we may reduce their search range for potential misbehaviours, 
such as SYN attacks or other DDOS attacks. 

The remainder of the paper is organised as follows. In Section~\ref{sec:rw}, we present a brief overview of the related work in disaster recovery and the use of SDN in this particular context. We then introduce our fault-tolerant distributed key-value store in Section~\ref{sec:storage}. In Section~\ref{sec:architecture}, we present our SDE architecture, its emulation and the performance results we obtained. 
Section~\ref{sec:discussion} discusses the opportunities and challenges of our solution. 
Finally,Ê Section~\ref{sec:conclusion} concludes. 

\section{Related Work}
\label{sec:rw}

In this section, we present an overview of the storage literature with a particular focus on disaster recovery and how Software Defined Network has been used to cope with scalability and fault tolerance.
As far as we know, no SDN solution currently addresses disasters, mainly because existing work focus on privately owned networks or datacenter networks. 

\subsection{Disaster recovery and storage}
\label{ssec:dr}

The dramatic impact of disasters on the persistence of data has long motivated database research community to investigate how to efficiently cope with disasters~\cite{GPH90}. 
This topic has regained attention in the last few years, with the ermergence of centralised cloud storage services within large datacenters around the globe~\cite{WLR+11}.

In the literature, two metrics usually characterise the persistence of data upon disaster. First, the Recovery Point Objective (RPO) represents the acceptable amount of data that can be lost, while the 
Recovery Time Objective (RTO) defines the amount of downtime that is permissible before the system recovers. 
Existing replication techniques usually minimise service latency overhead by executing asynchronously, they periodically copy the freshly stored 
data or a virtual machine image. Asynchronous replication was shown effective in practice to limit the bandwidth cost~\cite{PMF+02} even though it cannot achieve a nil RPO. 
Several companies already offer cloud storage solutions with a reasonably low, yet non-null, RPO.
In contrast with asynchronous replication, synchronous replication consists of duplicating the update request so that the two copies remain consistent~\cite{WLR+11}, hence promising to achieve a nil RPO.

Replication is the key to make data resilient to failures.
A distributed storage uses replication to guarantee that requests issued by clients get served despite the crash of a server. In the case of a disaster that affects a whole datacenter, 
it is important to geo-replicate data by copying the data across datacenters located in different regions of the globe.
Database solutions traditionally classify servers into a single primary and multiple backups. 
In such context, they exist two ways of recovering from disaster recovery~\cite{GPH90}. First, the client contacts the primary and the primary exchange messages with the backup(s) 
before the clients gets a response. To recover the data after a disaster at least one backup should be located in a different region from the primary, which makes this solution, 
called \emph{2-safety}, slow as the client request latency increases with the distance between regions. Second, the client contacts only the primary to get a faster response before 
any message exchange with the backups, an efficient alternative called \emph{1-safety} that unfortunately cannot guarantee recovery. 

Our solution falls between these two approaches. It requires the client to differentiate regular traffic that uses a 1-safety-like approach from the critical traffic 
that uses a 2-safety-like approach. The critical traffic typically consists of update 
requests that aim at storing critical data, like financial transactions, that should not be lost despite disasters. By contrast, the regular traffic consists of read-only requests
the server does not have to keep track of and normal update requests that are simply guaranteed to survive isolated failures but should complete rapidly. Given that latencies 
to datacenters vary significantly\footnote{RTT between Oregon and Singapore Amazon EC2 regions is $10\times$ the RTT from Oregon to California Amazon EC2 regions as 
reported at \texttt{http://www.bailis.org/blog/communication-costs-in-real-world-networks}}, such a distinction is important to only experience high latency during critical requests.

An alternative to primary/backups is to group servers by majorities, by simply redirecting any request issued by a client to a quorum of the servers. 
This strategy experiences a longer delay for read-only requests as a client must always wait for the participation of a number of servers that is linear in the total number of servers
before getting an acknowledgement from the storage service. This is typically the technique used to synchronise remote servers based on the Paxos consensus protocol.
%
Instead of considering majorities, an alternative is to exploit \emph{quorums}, or mutually interesting sets, of servers that indicate the minimum amount of 
servers where the data should be replicated~\cite{RCO+12}. While our solution is similar to the combination of 1-safety and 2-safety approaches, it actually balances the load on 
quorums of servers rather than using the primary/backups approach. 

As far as we know, our approach is the first one to reach a $30s$ RTO and a nil RPO, it lets the network recover rapidly from a 
disaster to make it almost transparent to the storage application.

\subsection{Software defined networks}
\label{ssec:sdn}


Software defined network (SDN) has been popularised by the decoupling of control functions and the data processing and forwarding functions, which makes data forwarding rules remotely controllable. 
OpenFlow is one way of defining the control over a collection of input/output ports (the switch) and a flow table\footnote{http://archive.openflow.org} that was successfully deployed in the GENI network in the US, the JGN2plus network in Japan and the Fed4FIRE project in Europe.

SDN can be used to adapt networks upon failures. This is especially appealing for datacenters where most of the network-dependent cloud services tend to be gathered.
This has led to a large body of work on the topic of datacenter networking~\cite{TCK+09}.
At a lower level, SDN offers inherent fault tolerance capabilities. For example, it allows a controller to use link-layer discovery protocol (LLDP) to detect link failures in a timely 
fashion, and to adjust in response to such a detection the behavior of the switches it controls. While this early approach may not be adequate for scalability~\cite{KBK+12}, recent 
versions of OpenFlow even allow the switch to react in face of failures without the continuous help from the controller~\cite{RCG+12}.

The use of SDN-enabled routers is also appealing for offering a cost-effective alternative to 
feature rich routers by moving the complexity from the switch itself 
to a separate controller, hence allowing an application-sensitive traffic without embedding the associated complexity at the switch level. Our solution uses a similar concept to 
distinguish critical traffic from non-critical traffic and duplicate the flow towards remote datacenters.

While SDN can be used as a mean for fault-tolerance, a controller is not immune to failures. It is clear that replication of controllers is not only necessary to scale to a large network
but also to allow some controllers to take over the role of failed controllers. The use of multiple controllers was already shown as effective in reducing the fault-tolerance of a 
particular SDN using NOX controllers and the Mininet virtualised environment~\cite{KST+12}. 
However, disasters typically impact a large region and may potentially affect a large amount 
of geo-localized network infrastructure equipments, including controllers.

Our solution also guarantees scalability and tolerance to isolated failures by multiplying controllers to forward traffic.
The novelty of our SDE approach lies, however, in distributing controllers at the edge of the network to even tolerate disasters affecting potentially an entire privately owned or 
datacenter network.

\section{The Key-Value Store Use-Case}
\label{sec:storage}
While appealing, the replication needed for fault-tolerance also raises problems related to the consistency of data, by potentially introducing out-of-date replicas or having 
different versions of the same data hosted at geographically distant datacenters. 
Hence communication must occur between servers to ensure that the new value of a data updated by a client gets propagated to multiple servers. 
There are various forms of consistency provided by key-value stores whose strongest is strong consistency or atomicity~\cite{HW90}. Some guarantee eventual consistency tolerating transient inconsistencies while others, like Yahoo's PNUTS, provides another form of consistency, called ``timeline'', among replicas hosted at remote datacenters. 

\begin{figure}[htb]
  \begin{center}
    \includegraphics[width=0.6\columnwidth]{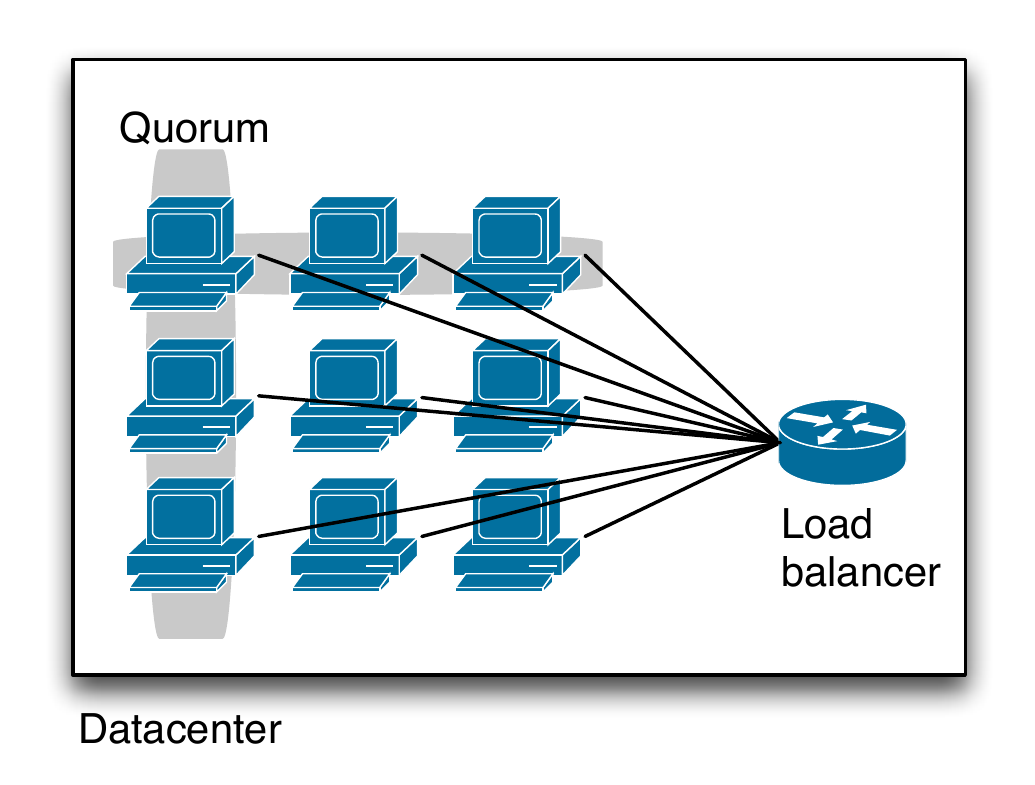}
    \caption{Quorum-based replication within a datacenter}
    \label{fig:datancenter}
  \end{center}
\end{figure}

To make sure of the uniqueness of the value of a data, two concurrent updates of the same data, say \texttt{put(k,v)} and \texttt{put(k,v')} requests, should be consistently ordered by all servers.
This can be achieved using one timestamp per value. This timestamp is computed based on the unique IP address of the issuing client and a global counter that together ``tag'' the version of a value, indicating  for example that \texttt{v'} is more up-to-date than \texttt{v}.
If needed, any server can potentially retrieve and compare the tag associated to a value to totally order the versions, hence deciding uniformly upon the most recent value of a data.

Another consistency requirement is that an update issued after a previous update on a same key has completed should always have a value tagged with a larger version.
This requires that each request, whether it be a read-only, like a \texttt{get}, or an update, like a \texttt{put}, starts by asking a majority of servers about the current highest version before proceeding
with propagating the most up-to-date value with its most up-to-date version.
Primary/backups, majorities and quorums are all sets of servers sharing similar intersection properties generalised by the notion of \emph{biquorum system}, 
corresponding to two classes of server sets such that any set of the first class, say \emph{read-qorums}, always intersect all sets of the second class, say \emph{write-quorums}.
One example is the rows and columns of servers as depicted in Figure~\ref{fig:datancenter}.
Using biquorum system, the server receiving the client \texttt{get} or \texttt{put} request from the client, first obtains the highest value-version pair from a read-quorum (this is the the value with the highest version among the ones received from each server of the quorum) before propagating the highest value-version pair to a write-quorum~\cite{CGGMS09} (making sure that all servers of this quorum stores it locally), such that any read-quorum has at least one server in common with any write-quorum.

To tolerate general failures our approach is twofold. First, we replicate all data within a datacenter where communication is low. 
This guarantees that the data persists despite isolated failures. Second, we replicate the critical traffic to a dacacenter towards a second datacenter located in a different region.
This is the responsibility of the client to distinguish normal from critical data as cross-regions traffic induces significant delays. 
Provided that both datacenters share a common initial state, say as given by some common virtual machine image, 
replicating the critical traffic across datacenters guarantees that critical data are not lost upon geo-localized disaster, even if an entire datacenter goes down. 
We present the architecture to redirect and replicate traffic across distant datacenters in the next section.

\section{Redirection of Traffic with SDE}
\label{sec:architecture}
In this section we present a simple yet scalable architecture using SDN technologies at both end of the network to implement 
a  distributed disaster-tolerant key-value store.
In particular we propose to deploy SDN technology as close as possible to the client in order to transparently modify the service destination and thus achieving our goal of nil RPO  and $30s$ RTO. 

\begin{figure}[htb]
  \begin{center}
    \includegraphics[width=1\columnwidth]{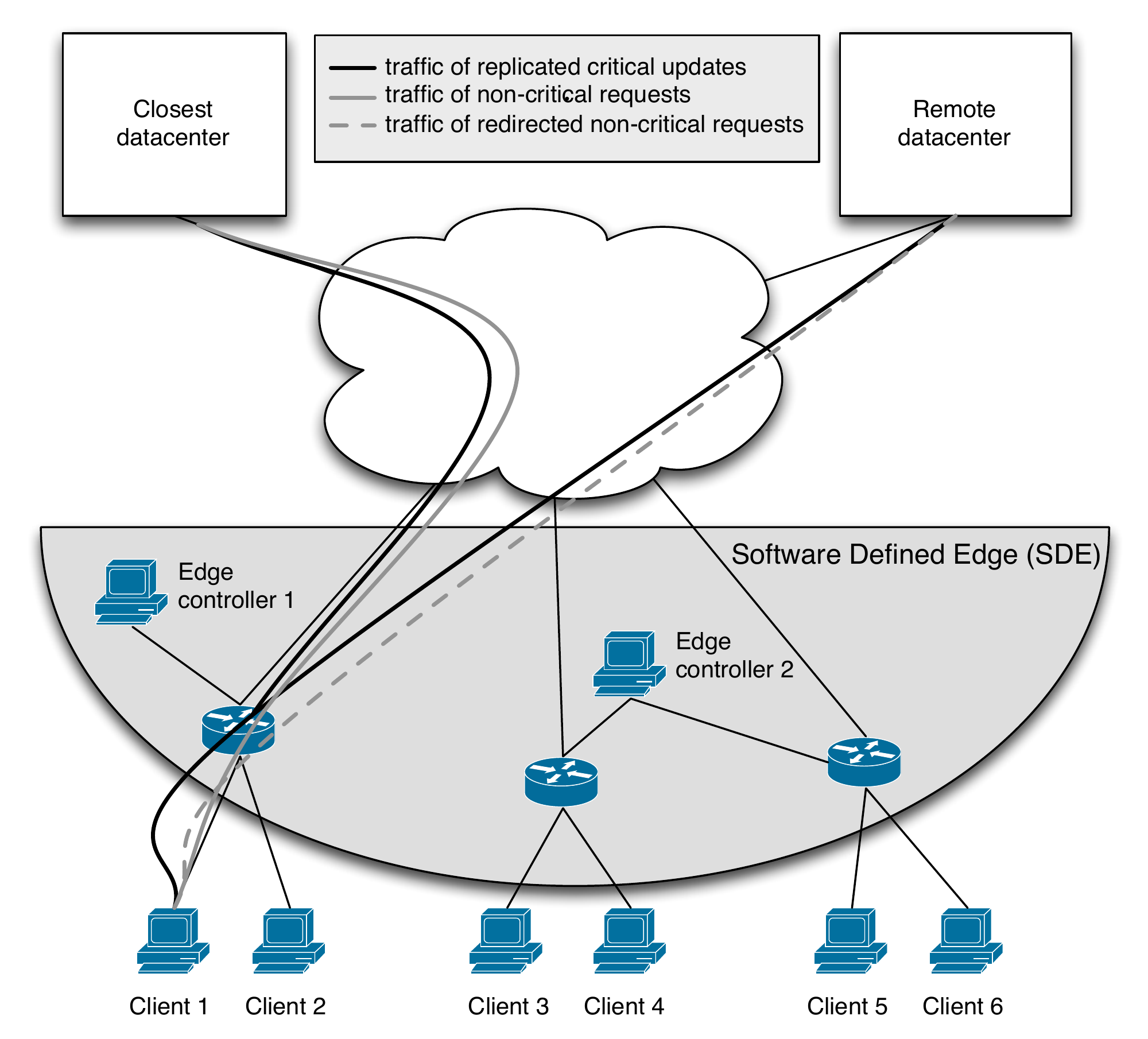}
    \caption{The Software Defined Edge (SDE) replicates critical traffic and redirects (non-critical) traffic upon disaster}
    \label{fig:architecture}
  \end{center}
\end{figure}

\subsection{Architecture of the Software Defined Edge}

The overall architecture is presented in \figurename~\ref{fig:architecture} with a simplified case of a single client connected to the key-value store service deployed over two datacenters. This client can perform two kinds of actions, namely critical update requests, and non-critical requests to the key-value store. In \figurename~\ref{fig:architecture}, these two actions are symbolised by respectively the black and grey connections. Dashed lines depict the traffic automatically redirected by our architecture upon disaster towards the remote backup datacenter.

In order to deploy our presented architecture, the edge routers on the client side are SDN-capable and connected to an SDN controller. This controller is responsible for the overall disaster resilience and we will detailed its behaviour in the remainder of this section. Furthermore, we also deploy a controller at the entrance of the datacenter in order to balance the load and increase the overall throughput.

%

For non-critical requests like \texttt{get} requests, we propose to use a rapid redirection mechanism. This mechanism is triggered only when the switch controller has detected that the primary datacenter is disconnected from the network by observing that clients cannot connect to their local datacenter. Once the redirection is decided and the controller sends a message to the switches it controls, these switches change the IP headers within the corresponding TCP flow to redirect it to the secondary datacenter and change back this header on return packets from the datacenter. From the client point of view the connection to the datacenter never changes and appears identical even in the occurrence of a disaster in the region of the primary service provider. We detail this mechanism in the Section~\ref{sec:results}. 

In order to decide when to start the redirection, we propose a mechanism inside the edge controller for the detection of potential disaster in the region of the main datacenter. This algorithm takes information from the various edge switches while the decision is centrally taken by the SDN controller responsible of these switches. This centralised algorithm is presented in Algorithm~\ref{alg:detection} and aims at minimising the RTO depending on the number of services currently deployed.

\begin{algorithm}[H]
\SetAlgoLined
 \KwData{Client flow to service}
 \KwResult{Possible detection of disaster }
 detection\;
 
  \eIf{packet duplicate packet from client}{
   possibleDisaster ++ \;
    \If{possibleDisaster $\geq$ Threshold}{
     Disaster mode = True\;
     \For{every connections affected by disaster}{
         send RST packet\;
         }
     }
   }
   {\If{ packet from Service}{
   
   possibleDisaster = 0\;
   }
   }
 
 \caption{Detection of disaster }
 \label{alg:detection}
\end{algorithm}

\subsection{Fast Redirection upon Network Failures}
\label{sec:results}

The idea is to place the control plane at the edge to cope with geo-localized disasters.
To this end, we use Mininet and the Python-based POX controller to evaluate the performance of our geo-replicated key-value store in the face of disaster.

\begin{figure}
  \begin{center}
    \includegraphics[width=.5\columnwidth]{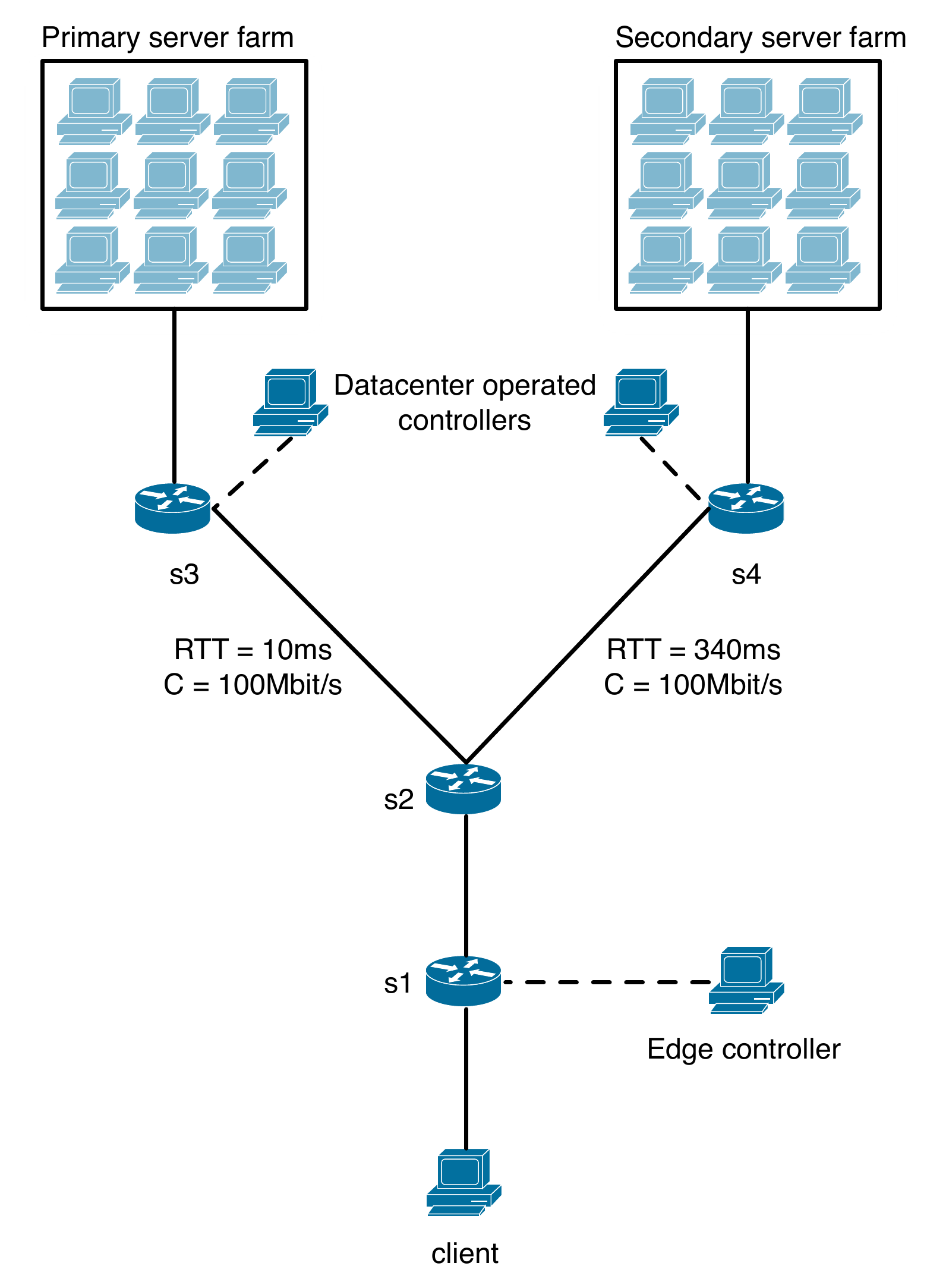}
    \caption{Experimental topology deployed in Mininet to represent a client accessing a local datacenter in Sydney (left) and a remote datacenter in Ireland (right)}
    \label{fig:topo}
  \end{center}
\end{figure}

In this section, we present preliminary results of the proposed architecture depicted in \figurename~\ref{fig:architecture} obtained using the mininet~\cite{HHJLM12} software and using POX controller. As a result we have deployed our solution over the topology presented in \figurename~\ref{fig:topo}. 
This topology depicts a local Amazon EC2 datacenter in Sydney and a remote Amazon EC2 datacenter located in Ireland: 
the link from the
switch $s_2$ to $s_3$ has a one-way delay of $5$ms and a capacity of $100$Mbit/s whereas the link from $s_2$ to $s_4$ has a one-delay of $170$ms and a capacity of $100$Mbit/s as the RTT between Sydney and Ireland Amazon EC2 datacenters is known to be about $340$ms.
Inside each datacenter, we have configured the key-value store over a quorum system of nine servers as depicted in Section~\ref{sec:storage}. Furthermore, both switches $s_3$ and $s_4$ act
as load balancers inside each datacenter.

Our experiment scenario is as follows. At time $t=0$, one of the client is starting reading for the local datacenter as fast as possible and thus filling the pipe between $s_2$ and $s_3$. At time $t=200$, we emulate the disaster by putting down the link between $s_2$ and $s_3$. Once the disaster occurs, the TCP connection sends retransmission packets and therefore notifies the controller that a possible disaster occurs in the network. As we fixed the detection threshold described in Algorithm~\ref{alg:detection} to 5 packets, the waiting period varies from $15$ to $25$ seconds. After this inactive period, the client reconnects to the service with the same IP address but is directed to the second datacenter.

\begin{figure}
  \begin{center}
   \subfloat[RTT]{\label{fig:rtt}{\includegraphics[width=.5\columnwidth]{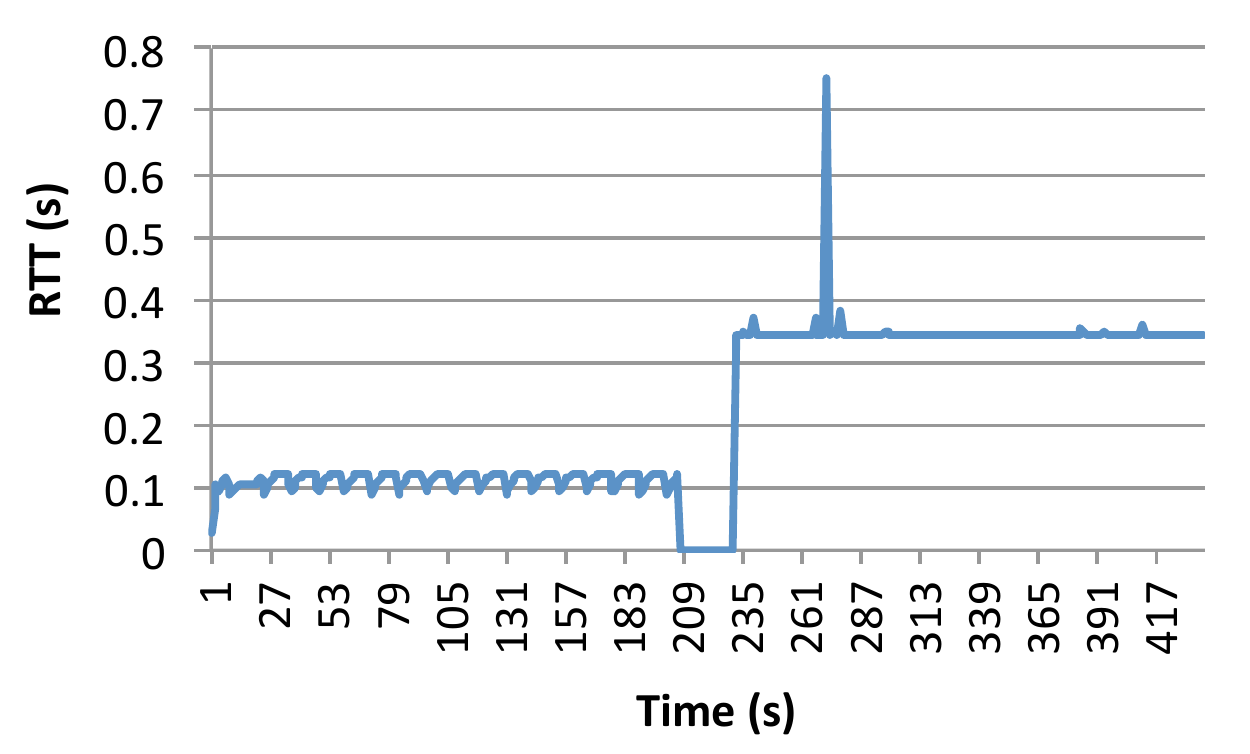}}}
   \subfloat[Throughput ]{\label{fig:throughput}{\includegraphics[width=.5\columnwidth]{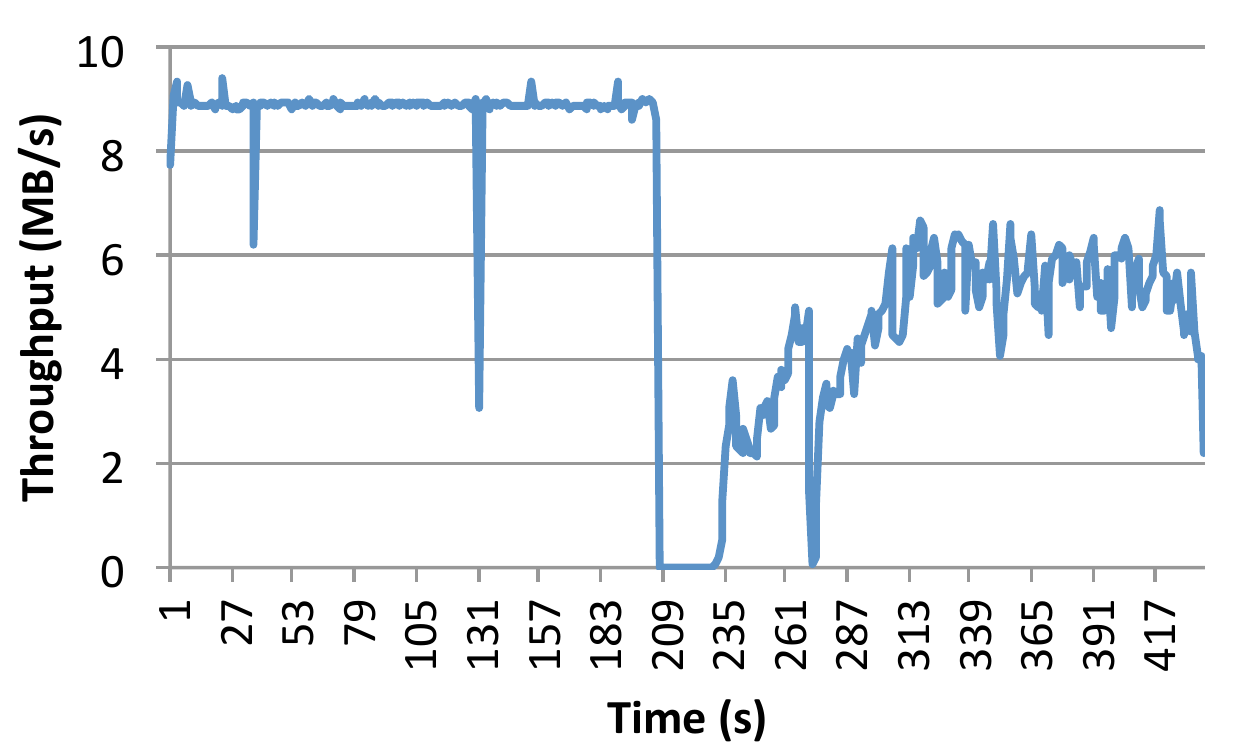}}}
    \caption{Impact of the disaster on service RTT and throughput of the key-value store}
    \label{fig:results}
  \end{center}
\end{figure}

In \figurename~\ref{fig:results}, we present our preliminary results extracted from the captured trace at the client side. The \figurename~\ref{fig:rtt} shows the round trip time computed by the TCP connection. In particular, we can see that during the first $200$s the flow suffered an RTT of around $100$ms, which is induced by the artificial delay we have introduced and the queueing delay when emulating the link capacity of $100$Mbit/s. Then, during the final $200$s period, the flow experiences the  minimal RTT of $340$ms necessary to reach the remote datacenter. \figurename~\ref{fig:throughput} presents the throughput experienced by the TCP flow. As expected, when the delay is short, the TCP connection is able to fill the pipe when accessing the local datacenter while a larger RTT makes it impossible for TCP to fully utilise the pipe. Using this SDE technique, we observed empirically an RTO lower than 30s.  

\subsection{Replication of Traffic with SDE}
\label{sec:dupli}

\begin{figure}
  \begin{center}
    \includegraphics[width=0.9\columnwidth]{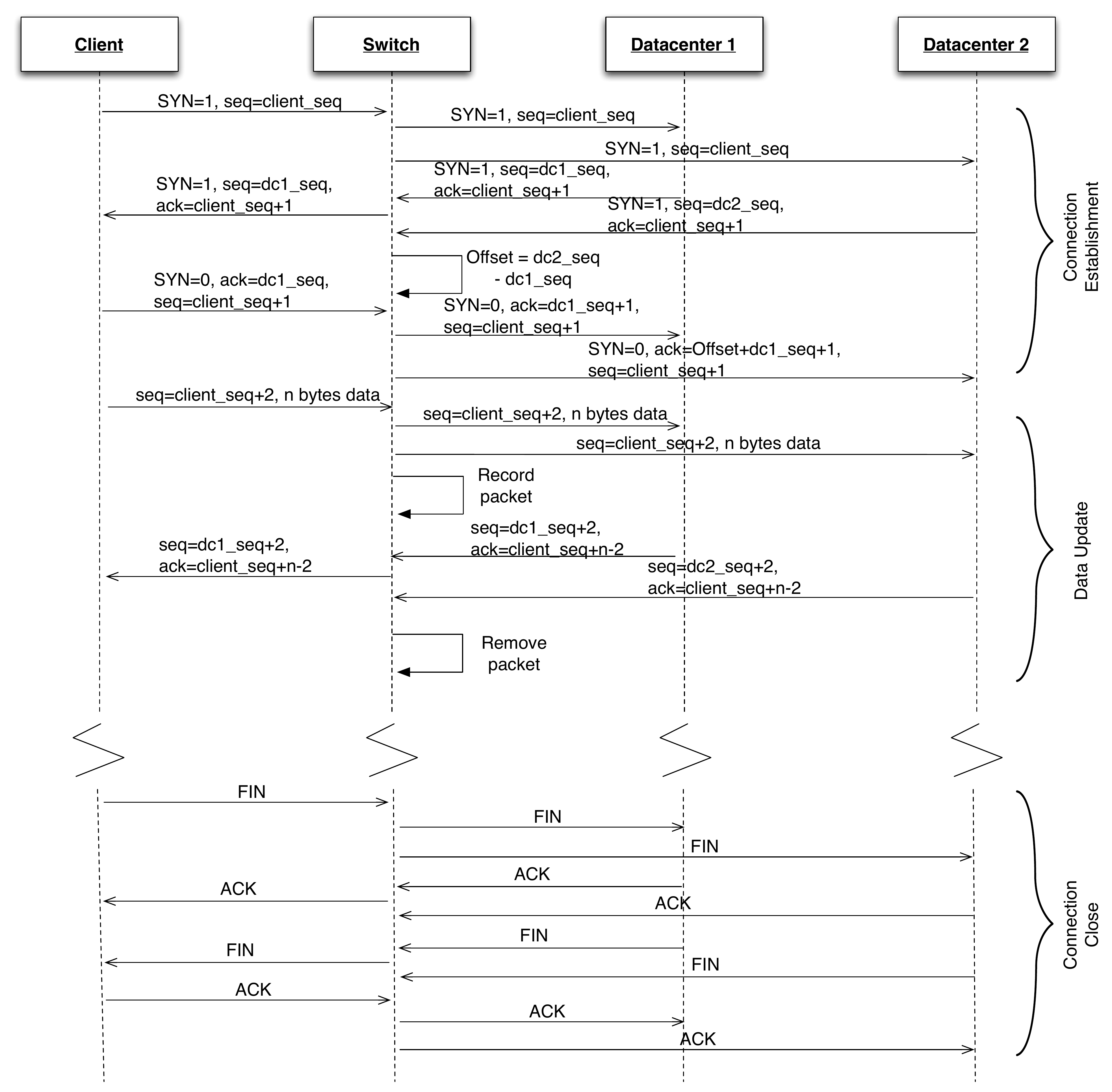}
    \caption{Mechanism for inline TCP duplication}
    \label{fig:dupli}
  \end{center}
\end{figure}

To ensure persistence of critical data (nil RPO), we extend the redirection mechanism used in the former case by proposing a duplication (or replication) of the TCP flow in the non-disaster period to achieving a 30s RTO. This mechanism is illustrated in \figurename~\ref{fig:dupli} where the switch will translate the data from the client to the second datacenter in order to mimic a connection to the client. 
This replication is made possible by extending the flow table in the switch. In particular, at the establishment of the TCP connection the switch duplicates the SYN packet to both datacenters, while creating an entry in the flow table, then when the first ACK is coming back, the switch elects the datacenter as the master datacenter and keeps a trace of its acknowledged sequence number. Once the second and more distant datacenter sends back the ACK packet, the switch elects this datacenter as the slave and computes the offset in the acknowledged sequence number. This offset is used in the on-going connection to make the slave server believes it is connected to the client. This process is illustrated in the upper part of the sequence diagram of \figurename~\ref{fig:dupli}.

Once the connection is established between the two servers and the client, the latter can start writing in the key-value store. In this case, the switch duplicates the TCP stream using the offset value as depicted in \figurename~\ref{fig:dupli}. As the connection to the second datacenter is longer than to the first datacenter, we need to copy non-acknowledged packets from the slave datacenter in the switch or in the SDN controller. 
While this imposes a constraint on the storage capability of the switch, we can reasonably expect that the number of flows as well as the size of these flows will not overflow the switch memory given that our solution is deployed at the edge of the network.
Using this copy of non-acknowledged packets, we can retransmit them in case of losses in the network.

Finally, we close the client connection to the datacenter in a similar manner as the connection establishment as shown in \figurename~\ref{fig:dupli}.

%
%
%


\section{Challenges and Opportunities}
\label{sec:discussion}

We illustrated how SDE can be used to mitigate disasters. Our proposal faces a number of challenges but also creates new economically viable opportunities. We identified three main challenges  as well as three main opportunities and we discuss both aspects in the remainder of this section. 

\begin{itemize}
\item The first challenge is related to the technical aspect of TCP flow duplication as we mentioned briefly in Section~\ref{sec:dupli}. Indeed, TCP transmission provides end-to-end reliability that naturally extends to the client to master datacenter connection but the replicated slave connection is beyond the reach of the client. The switch should therefore take care of the retransmission of lost packets based on acknowledgements from the slaves. In the previous section, we introduced an extension to the flow table to perform this action but further investigation is needed to decide when retransmissions are needed. A few options are available, such as timer-based triggers or using a threshold in the number of 
unacknowledged packets in the buffer. Another issue with TCP duplication concerns the asymmetry of the paths to the datacenters and its impacts on the sending rate. 
\item 
Second, the proposed architecture relies on its deployment at the edge of the network. However, it is unlikely that the datacenter operator would be the same as the edge ISP. Hence, the deployment requires collaboration between these two entities. This is a well-known issue for Quality of Service as some argued for the the establishment of an API to delegate some control over the edge switches to third parties~\cite{SMG+13}. 
\item
Finally, we demonstrated our proposal over a simple key-value store. Extension to more complex data storage backends might require additional work.
\end{itemize}

Notwithstanding these challenges, our proposed architecture presents several opportunities. 
\begin{itemize}
\item First, we have illustrated how our proposal achieves nil RPO while approaching a nil RTO. This opens a new business opportunity as this feature could be integrated in today's Service Level Agreements as a new premium option. 
\item 
Second, while deployment requires collaboration between ISPs and cloud-based service providers, the use of SDN technology greatly reduces its cost. Indeed, SDN-enabled switches are already available off the shelf and usually cost orders of magnitude less than more complex equipments. 
\item 
Finally, by proactively redirecting traffic to online datacenter upon disaster, our proposal would reduce the number of legitimate but unsuccessful connection establishments. This in turn would simplify the task of the edge ISP forensic department as a large number of failed connection establishments due to a disaster would have triggered DDoS investigation procedures. Furthermore, we could imagine that the redirection information would be shared with the ISP, which could therefore take the necessary steps to reconfigure their network faster (\textit{e.g.}, change BGP announces). 
\end{itemize}



\section{Concluding Remarks}
\label{sec:conclusion}
SDN is becoming a mature technology to adapt the control plane within few seconds by dynamically mapping the distribution of switches to controllers~\cite{LWH12}.
This reactivity promises to facilitate application fault-tolerance through highly responsive network failure detection. We have confirmed this idea through the use of a simple 
cloud key-value store application that can recover from continent-wide disaster almost transparently from the end-user point-of-view.
Although inter-datacenter SDN is well known to adequately leverage the application information for traffic differentiation, we have shown somehow a reverse ability for 
distributed applications to leverage SDN edge information about fault detection to effectively bypass disasters.

\bibliographystyle{plain}
\bibliography{bib}

\begin{thebibliography}{10}

\bibitem{CGGMS09}
Gregory Chockler, Seth Gilbert, Vincent Gramoli, Peter~M. Musial, and
  Alexander~A. Shvartsman.
\newblock Reconfigurable distributed storage for dynamic networks.
\newblock {\em Journal of Parallel and Distributed Computing}, 69(1):100--116,
  Jan 2009.

\bibitem{GPH90}
Hector Garcia-Molina, Christos~A. Polyzois, and Robert~B. Hagmann.
\newblock Two epoch algorithms for disaster recovery.
\newblock In {\em Proceedings of the 16th International Conference on Very
  Large Data Bases}, VLDB '90, pages 222--230, San Francisco, CA, USA, 1990.
  Morgan Kaufmann Publishers Inc.

\bibitem{HHJLM12}
Nikhil Handigol, Brandon Heller, Vimal Jeyakumar, Bob Lantz, and Nick McKeown.
\newblock Reproducible network experiments using container-based emulation.
\newblock In {\em CoNEXT}, 2012.

\bibitem{HW90}
Maurice Herlihy and Jeannette Wing.
\newblock Linearizability: a correctness condition for concurrent objects.
\newblock {\em ACM Trans. Program. Lang. Syst.}, 12(3), 1990.

\bibitem{KBK+12}
James Kempf, Elisa Bellagamba, Andr{\'a}s Kern, David Jocha, Attila Tak{\'a}cs,
  and Pontus Sk{\"o}ldstr{\"o}m.
\newblock Scalable fault management for openflow.
\newblock In {\em ICC}, pages 6606--6610, 2012.

\bibitem{KST+12}
Hyojoon Kim, J.R. Santos, Y.~Turner, M.~Schlansker, J.~Tourrilhes, and
  N.~Feamster.
\newblock Coronet: Fault tolerance for software defined networks.
\newblock In {\em Proceedings of the 20th IEEE International Conference on
  Network Protocols}, ICNP, pages 1--2, 2012.

\bibitem{LWH12}
Dan Levin, Andreas Wundsam, Brandon Heller, Nikhil Handigol, and Anja Feldmann.
\newblock Logically centralized?: State distribution trade-offs in software
  defined networks.
\newblock In {\em HotSDN}, pages 1--6, 2012.

\bibitem{MAB+08}
Nick McKeown, Tom Anderson, Hari Balakrishnan, Guru Parulkar, Larry Peterson,
  Jennifer Rexford, Scott Shenker, and Jonathan Turner.
\newblock Openflow: Enabling innovation in campus networks.
\newblock {\em SIGCOMM Comput. Commun. Rev.}, 38(2):69--74, March 2008.

\bibitem{PMF+02}
R.~Hugo Patterson, Stephen Manley, Mike Federwisch, Dave Hitz, Steve Kleiman,
  and Shane Owara.
\newblock Snap{M}irror: File-system-based asynchronous mirroring for disaster
  recovery.
\newblock In {\em Proceedings of the 1st USENIX Conference on File and Storage
  Technologies}, FAST '02, Berkeley, CA, USA, 2002. USENIX Association.

\bibitem{RCO+12}
Shriram Rajagopalan, Brendan Cully, Ryan O'Connor, and Andrew Warfield.
\newblock Second{S}ite: Disaster tolerance as a service.
\newblock In {\em Proceedings of the 8th ACM SIGPLAN/SIGOPS Conference on
  Virtual Execution Environments}, VEE '12, pages 97--108, 2012.

\bibitem{RCG+12}
Mark Reitblatt, Marco Canini, Arjun Guha, and Nate Foster.
\newblock Fat{T}ire: Declarative fault tolerance for software-defined networks.
\newblock In {\em HotSDN}, pages 109--114, 2013.

\bibitem{SMG+13}
Vijay Sivaraman, Tim Moors, Hassan~Habibi Gharakheili, Dennis Ong, John
  Matthews, and Craig Russell.
\newblock Virtualizing the access network via open {API}s.
\newblock In {\em CoNEXT}, pages 31--42, 2013.

\bibitem{TCK+09}
Arsalan Tavakoli, Martin Casado, Teemu Koponen, and Scott Shenker.
\newblock Applying {NOX} to the datacenter.
\newblock In {\em 8th ACM Workshop on Hot Topics in Networks}, HotNets. ACM
  SIGCOMM, 2009.

\bibitem{Vah13}
Amin Vahdat.
\newblock Scale and programmability in {G}oogle's software defined data center
  {WAN}.
\newblock In {\em ACM Symposium on Cloud Computing}, SoCC '13, 2013.
\newblock Personal communication.

\bibitem{WLR+11}
Timothy Wood, H.~Andr{\'e}s Lagar-Cavilla, K.~K. Ramakrishnan, Prashant Shenoy,
  and Jacobus Van~der Merwe.
\newblock Pipe{C}loud: Using causality to overcome speed-of-light delays in
  cloud-based disaster recovery.
\newblock In {\em ACM Symposium on Cloud Computing}, SoCC '11, pages
  17:1--17:13, 2011.

\end{thebibliography}
\end{document}